# Non-Ideal Absorbers in Hot-Carrier and Hot-Absorber Solar Cells: Losses and Heat Recycling

Inés Durán[1], Simon Svatek[1], Irene Artacho[1], Elisa Antolín[1], Antonio Martí[1*]

[1]Instituto de Energía Solar, Universidad Politécnica de Madrid, Avenida Complutense, 30, 28040, Madrid, Spain

* Corresponding author: antonio.marti@upm.es

**Abstract**

The hot carrier solar cell exhibits a photovoltaic conversion efficiency limit of 85.4%. Among the idealizations necessary to achieve this efficiency, it has traditionally been considered essential that the electrons absorbing sunlight do not end up transferring the captured energy to the crystal lattice. In this work, we argue that such a requirement is not necessary if the absorber is thermally insulated. In our analysis, we also consider the possibility of ohmic losses in the absorber due to finite electrical conductivity. We also demonstrate that such losses, delivered as heat, can be partially recycled into electrical work with an efficiency that is given by the Carnot factor. The high efficiencies reported here are thermodynamic limits obtained with ideal energy-selective contacts and negligible parasitic contact conductance, not projections of practical device efficiency. Our work includes a sensitivity analysis for finite contact-mediated heat leakage, and the results are presented as design targets.

## 1. Introduction

The original hot carrier solar (HCSC) concept was proposed in 1982 by Ross and Nozik [1]. The basic idea consists in extracting electrons from an absorber of sunlight and before the electrons thermalize (hence their description as "hot carriers") they are extracted through energy-selective contacts (ESCs). The model was later refined by P. Würfel [2]: he argued that a hot electron population in the absorber is better described by a single electrochemical potential, instead of two, one for electrons and one for holes, and he clarified the role of the ESCs as the elements in charge of reversibly. They reduce the electron temperature from their initial hot state to ambient temperature and thereby raise their electro-chemical potential.

The HCSC can thus be visualized as consisting of two elements: the absorber (that absorbs photons from the sun, captures their energy and raises the temperature of the electrons of the absorber to $T_0$) and the ESCs. The ESCs allow only electrons with energy $\epsilon_H$ to be transported, for the case of the high-energy selective contact (H-ESC) and $\epsilon_L$, for the case of the low-energy selective contact (L-ESC). During the transport, the ESCs changes reversibly the temperature and electrochemical potential of the electrons (from $\mu_0$ and $T_0$ to $\mu_{H,I}$ and $T_T$ at the H-ESC, and from $\mu_{L,I}$ and $T_T$ back to $\mu_0$ and $T_0$, at the L-ESC). This allows supplying to the external world, represented by the load in Fig. 1, useful work by electrons given by the electrochemical potential difference $\mu_{H,I} - \mu_{L,I}$. The existence of a single electrochemical potential, $\mu_0$, for the electrons in the absorber is a consequence of a HCSC model in which "energy conservation" is imposed as the only condition and electronic transitions between energy states in the absorber all occur in the same timescale. Imposing additional conditions, similar to the one stated in conventional solar cells based on semiconductors, in which, in addition, two-electrochemical potentials are assumed to exit (one for electrons and one for holes) leads to the so called "particle conservation model" but it has been found that it leads to non-plausible results [3].

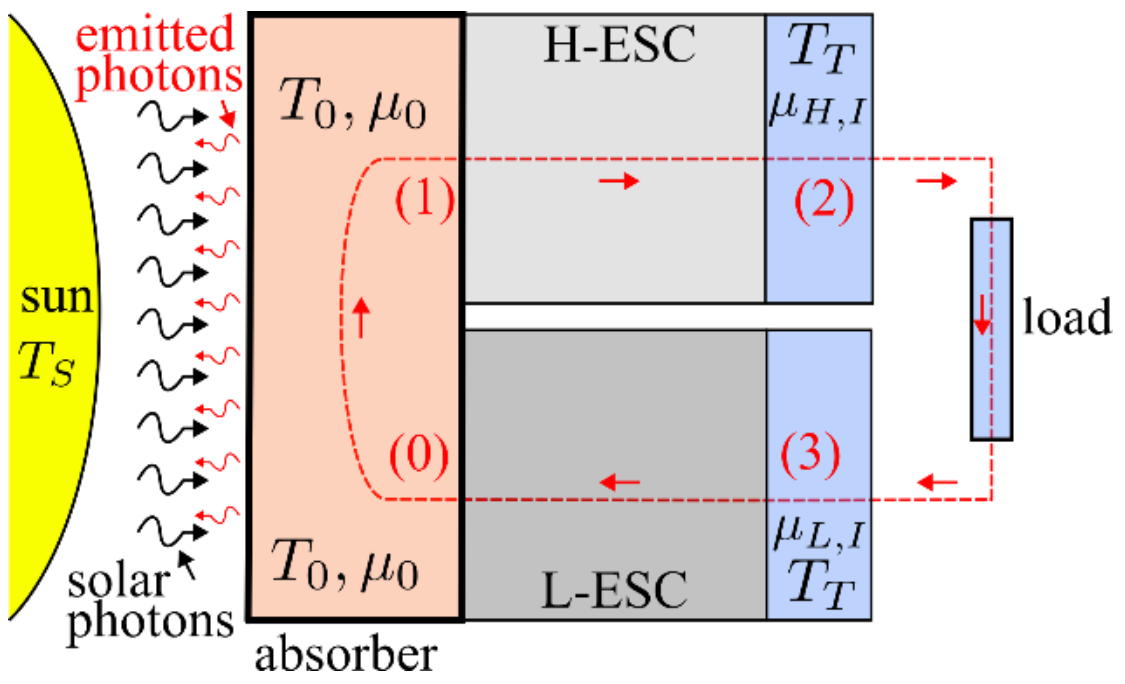

Fig. 1. Thermodynamic scheme of an ideal HCSC showing its two main components: the absorber and the energy selective contacts (high-energy selective contact: H-ESC; low-energy selective contact: L-ESC). The red arrows and numbers represent the direction in which electrons circulate when delivering positive work to the load. $T_S$, $T_0$ and $T_T$ represent the temperature of the sun, of the electrons in the absorber and the contact terminals respectively to which the load is attached. The temperature of the terminals, $T_T$, is assumed to be the same one as the ambient temperature which acts as cold reservoir. $\mu_0$ is the electrochemical potential of the electrons in the absorber; $\mu_{H,I}$ and $\mu_{L,I}$ correspond to the electrochemical potential of the electrons at the cold end of the H-ESC and L-ESC, respectively.

The interest in the HCSC stems from the fact that it offers a much higher theoretical power conversion efficiency limit, 85% for maximum concentrated sunlight [2], which is twice the Shockley-Queisser limit for than conventional single gap solar cells [4,5]. Part of this high limiting efficiency potential arises from the fact that, as mentioned, ideal ESCs are assumed to transport electron energy isentropically. In Ref. [6], the impact of non-ideal selective contacts in the performance of the HCSC was studied concluding, for example, that ideal ESCs behave like thermoelectrics that can demand ZT figures of merit higher than 8 to preserve HCSC efficiency above 84 % under concentrated light. [7] For context, state-of-the-art thermoelectric materials typically reach $ZT \sim 1$ in established systems [8,9] and peak $ZT \sim 2$–3 in leading research materials [10] at optimized temperatures; thus $ZT > 8$ highlights that near-isentropic energy-selective transport is an extremely stringent requirement. In this work, this study will be continued by exploring the impact on the performance of HCSCs of irreversible electron transport in the absorber. In this framework we will show, for example, how the HCSC is capable of recycling into electrical work part of the heat that is dissipated in the absorber due to the finite value of the electrical conductivity. After reviewing the formalism of the operation of the ideal HCSC in section 2 according to the energy conservation model and describing, in section 3, the model for the case of the absorber characterized by non-ideal electron transport, in section 4 we will define the hot absorber solar cell (HASC). This consists of a modification of the HCSC concept in which the common restriction for electron-phonon interaction in the absorber is lifted. In section 5 we will develop a model for the HCSC and the HASC that includes radiative losses from the perimeter and finite electrical conductivity in the absorber. This model is exploited in section 6 to analyze the impact in the efficiency of photon radiative emission through the perimeter, of finite electrical conductivity in the absorber, of heat recycling from the absorber, the energy difference of the ESCs, and finite contact-mediated heat leakage.

## 2. Review of the ideal HCSC energy conservation model

In the ideal HCSC absorber and under the energy conservation model, the energy captured from solar photons can only exit from the absorber either by being transported by electrons through the ESCs or by the photons emitted from the absorber. Hence, in the ideal HCSC, energy conservation states that:

$$P_a - P_e = \frac{I}{e}(\epsilon_H - \epsilon_L) \tag{1}$$

where $P_a$ is the energy per unit of time absorbed from the sun, $P_e$ is the energy per unit of time corresponding to the emission of photons, $e$ is the elementary charge and $I$ is the electrical current extracted from the cell. This is the "energy conservation model" mentioned in our introduction. To calculate the maximum efficiency under a common reference, usually the sun is modelled as a black body at temperature $T_S$ and the absorber as a blackbody at temperature $T_0$ what leads, when maximum solar concentration is assumed, to:

$$P_a = \hat{\sigma} T_S^4 A_S \tag{2}$$

$$P_e = \hat{\sigma} T_0^4 A_S \tag{3}$$

where $\hat{\sigma}$ is the Stefan-Boltzmann constant and $A_S$ is the absorber area through which photons are received from the sun and emitted from the absorber. When electrons in the ideal HCSC absorber absorb sunlight and capture its energy, they reach the temperature $T_0$ and electrochemical potential $\mu_0$. The temperature $T_0$ depends on the value of the external load attached to the HCSC since its value impacts the electrical power extracted from the cell being, for example, $T_0 = T_T = 300$ K (ambient temperature) at short-circuit and $T_0 = T_S = 6000$ K (sun temperature) at open-circuit. [2,6,11,12]. $\mu_0$ does not change because it is taken as the electrochemical zero reference at all solar cell operation conditions.

Fig. 1 also illustrates the electron flow in an ideal HCSC along a closed path. Starting at point 1, electrons exit from the absorber and, since they are at equilibrium with the absorber at this point, they are also characterized by the temperature $T_0$ and electrochemical potential $\mu_0$. Electrons are then transported through the H-ESC from point 1 to 2. During this transport, electrons preserve their energy $\epsilon_H$ (which is also the energy that characterizes the energy of the H-ESC) but lower isentropically their temperature from $T_0$ to the contact temperature $T_T$ (which corresponds also to the ambient temperature) while they raise their electrochemical potential from $\mu_0$ to $\mu_{H,I}$ given by [2]:

$$\mu_{H,I} = \mu_0 + (\epsilon_H - \mu_0)\left(1 - \frac{T_T}{T_0}\right) \tag{4}$$

At the load, electrons lower their electro-chemical potential and exit from point 3 with the same temperature $T_T$ and electrochemical potential $\mu_{L,I}$ given by:

$$\mu_{L,I} = \mu_0 + (\epsilon_L - \mu_0)\left(1 - \frac{T_T}{T_0}\right) \tag{5}$$

From point 3 to point 0, electrons circulate through the L-ESC characterized by the energy $\epsilon_L$, preserving this energy during this transport, but they isentropically raise their temperature back from $T_T$ to $T_0$ and change their electrochemical potential from $\mu_{L,I}$ to $\mu_0$.

The electrochemical energy per electron released to the load, $e \cdot V,$ is therefore, in this ideal case, given by:

$$e \cdot V = \mu_{H,I} - \mu_{L,I} = (\epsilon_H - \epsilon_L)\left(1 - \frac{T_T}{T_0}\right) \tag{6}$$

where $V$ can be visualized as the voltage appearing at the terminals of the load (measured from the L-ESC to the H-ESC). When the electrical current $I$ circulates through the load, the electrical work obtained from the HCSC per unit of time, $\dot{W}_{id}$, in this ideal case is then given by:

$$\dot{W}_{id} = I \cdot V = \frac{I}{e}\left(\mu_{H,I} - \mu_{L,I}\right) = \frac{I}{e}(\epsilon_H - \epsilon_L)\left(1 - \frac{T_T}{T_0}\right) \tag{7}$$

By varying $T_0$, the current-voltage characteristic of the ideal HCSC can be obtained parametrically from Eqs. (6) and (1) and, from this characteristic, the maximum work $\dot{W}_{id,max}$. The maximum efficiency, $\eta$, is then defined as,

$$\eta = \frac{\dot{W}_{id,max}}{P_a} \tag{8}$$

from which the known 85 % limiting efficiency of the ideal HCSC is obtained.

It is also worth noticing that energy conservation per electron during transport between points 2 and 3 reads:

$$\epsilon_H - \epsilon_L = \mu_H - \mu_L + Q \tag{9}$$

where $Q$ is the net heat released at the contacts per electron transported. Using (6),

$$Q = (\epsilon_H - \epsilon_L)\left(\frac{T_T}{T_0}\right) = Q_H - Q_L \tag{10}$$

where, following Ref. [6],

$$Q_H = \epsilon_H \frac{T_T}{T_0} \tag{11}$$

can be interpreted as the heat released per electron at contact point 2 to the heat reservoir (the ambient, that keeps the extreme of the contacts at temperature $T_T$) and

$$Q_L = \epsilon_L \frac{T_T}{T_0} \tag{12}$$

as the heat reabsorbed at point contact 3 from the ambient (which can be negative if $\epsilon_L < 0$, that is, if $\epsilon_L < \mu_0$ since $\mu_0 = 0$ is being taken as the zero electrochemical potential reference). Further additional details, such as for example, the fact that when $\epsilon_L < \mu_0$, transport through the L-ESC can be regarded as hole transport and as electron transport when $\epsilon_L > \mu_0$ can also be found in Ref. [6].

This has been, so far, a condensed description of the operation of an ideal HCSC. The first cause of non-ideality arises from the contacts when electron transport through them becomes non-isentropic. However, as mentioned in the introduction, the impact of this non-ideality was already discussed by some of us in Ref. [6]. In this broader context we also note the heat-recovery solar cell (HERC) concept [13], which employs energy-selective transport to extract electrical work from an absorber that is hotter than the terminals by recovering generic heat flows (e.g., phonon-mediated or externally supplied heat). By contrast, in this work we retain the HCSC energy-conservation framework with ideal ESCs and focus on a thermally isolated ("hot absorber") scenario in which the only additional heat is the internally generated Joule heat arising from finite conductivity in the absorber. Accordingly, the calculations below adopt the ideal-contact boundary $g_{ESC} = 0$ while retaining selected-carrier energy transport and the current-coupled cold-side heat rejection in Eqs. (9)–(12). These rates are proportional to $I$; they are distinct from the independent parasitic ESC heat-leakage pathway introduced later. We then focus on the consequences of non-ideal electron transport through the absorber.

## 3. HCSC absorber with non-ideal electron transport

Electron transport in the absorber is necessary to take electrons from the L-ESC back to the H-ESC (that is, from point 0 to 1 in Fig. 1). When electrons circulate through the absorber, they raise their energy from $\epsilon_L$ to $\epsilon_H$, thanks to energy taken from photons and, in an ideal absorber, they are able to do so while they preserve their temperature $T_0$ and electrochemical potential $\mu_0$ during this transport.

However, if the absorber is assumed to operate at uniform temperature $T = T_0$, so that $\boldsymbol{\nabla} T = 0$, only a gradient in the electron electrochemical potential, $\mu$, can act as driving force to induce the necessary electron current-density, $\boldsymbol{j}_n$, that carries inside the absorber electrons from the L-ESC to the H-ESC. This implies that a constant value for the electron electrochemical potential in the absorber, such as $\mu_0$ when the ideal HCSC was described, cannot be assumed any more, unless infinite electron mobility, and therefore also infinite electrical conductivity, were assumed. Nevertheless, although a uniform electrochemical potential in the absorber cannot be assumed, still an electrochemical potential reference at a given single point, such as $\mu = \mu_0 = 0$ at point 0, can still be established (Fig. 2).

The electron current density $\boldsymbol{j}_n$ (measured as number of electrons per unit of time and unit of area) is related to the electron electrochemical potential gradient by [14]:

$$\boldsymbol{j}_n = -\frac{\sigma}{e^2} \boldsymbol{\nabla} \mu \tag{13}$$

where $\sigma$ is the absorber electrical conductivity.

In an ideal absorber, $\sigma \to \infty$, and the electron electrochemical potential, $\mu \equiv \mu_0$ becomes, as anticipated before, uniform inside the absorber since we must have $\boldsymbol{\nabla} \mu \to 0$ to preserve a finite value for $\boldsymbol{j}_n$. The entropy generation rate, $\dot{s}$, per unit of volume associated with electron transport is given by:

$$\dot{s} = -\frac{1}{T_0} \boldsymbol{\nabla} \mu \cdot \boldsymbol{j}_n = \frac{e^2 j_n^2}{T_0 \sigma} \tag{14}$$

where we have used Eq. (13). Notice that, consistently with the ideal transport concept, $\dot{s}$ also approaches zero when $\sigma \to \infty$ . However, for finite $\sigma$, a total entropy $\dot{S}$ is generated at the volume absorber $\Omega_0$ given by:

$$\dot{S} = \int_{\Omega_0} \dot{s} \, d\Omega = R I^2 / T_0 \tag{15}$$

The former equation is used to define the series resistance, $R$, associated to the electron transport within the absorber. This definition is the same one usually used in solar cells [15]. Its utility relies on the fact that, once it has been determined, it allows to calculate straight forward the power losses associated to this irreversibility for a given current.

However, since in the ideal HCSC, electrons are not allowed to interact with the phonons of the absorber, this additional irreversible entropy $\dot{S}$ generated in the absorber volume must be evacuated as heat elsewhere, which necessarily must be at the contacts, through points 2 and 3, since these are the only points in which the HCSC is in contact with the ambient reservoir. The additional heat released per unit of time, $\dot{Q}_a$, at these contact points is given then by:

$$\dot{Q}_a = T_T \dot{S} = RI^2 \frac{T_T}{T_0} \tag{16}$$

and must be subtracted from the ideal work in (7) to calculate the actual work per unit of time, $\dot{W}$, delivered to the load in this non-ideal situation:

$$\dot{W} = \frac{I}{e}\left(\mu_{H,I} - \mu_{L,I}\right) - RI^2 \frac{T_T}{T_0} \tag{17}$$

Hence, we observe that, while the power dissipated by a conventional resistor of value $R$ would be given by $RI^2$, in a HCSC this power loss is decreased by the factor $T_T/T_0$. To dig into the physical interpretation of this fact, we notice that $\dot{W}$ in Eq. (17) can also be written as:

$$\dot{W} = IV = \frac{I}{e}\left(\mu_{H,I} - \mu_{L,I}\right) - RI^2 \frac{T_T}{T_0} = \frac{I}{e}\left(\mu_{H,I} - eRI\frac{T_T}{T_0} - \mu_{L,I}\right) = \frac{I}{e}\left(\mu_{H,R} - \mu_{L,R}\right) \tag{18}$$

with

$$\mu_{H,R} = \{\mu_0 - eRI\} + (\epsilon_H - \{\mu_0 - eRI\})\left(1 - \frac{T_T}{T_0}\right) \tag{19}$$

$$\mu_{L,R} = \mu_{L,I} \tag{20}$$

and

$$V = \frac{1}{e}\left(\mu_{H,R} - \mu_{L,R}\right) \tag{21}$$

Comparing with Eq. (4), we see that the term in braces in Eq. (15), $\{\mu_0 - eRI\}$, can be interpreted as a reduction in the electrochemical potential at point 1 with respect to 0 by $eRI$ and justifies the insertion of a resistor between points 0 and 1 in Fig. 2 to model this impact on the HCSC equivalent electric circuit.

Hence, we can interpret the situation as follows. The ohmic losses in the absorber generate a power loss inside the absorber of $RI^2$. However, this power, since it is dissipated within the absorber, assists to keep the absorber hot and electrons within the absorber can reabsorb part of this power to produce work. However, because this power is not entropy free, it cannot be completely converted into useful work at the load. To accomplish the second law of thermodynamics, the extra entropy generated by the electron transport through the absorber must be released as heat at the contacts with the load (16). Therefore, the power $RI^2$ that is recycled into work thanks to the HCSC operation is:

$$\dot{W}_R = RI^2 - \dot{Q}_a = RI^2\left(1 - \frac{T_T}{T_0}\right) \tag{22}$$

which corresponds to the Carnot efficiency and is a logical consequence derived from the fact that electron transport through the ESCs has still been considered ideal.

The Joule-heat recycling described by Eqs. (13)–(22) requires $R\,I^2$ to be generated inside the hot absorber reservoir and retained there rather than immediately escaping through a bypass. The absorber is treated as spatially isothermal and internally equilibrated at $T_0$, the ESC conversion is ideal and reversible, the cold terminals remain at $T_T$, and nonradiative bypass leakage is negligible in the ideal limit.

These conditions distinguish the model from an ordinary photovoltaic series resistance. In an ordinary cell the dissipative element is effectively at or near the cold device temperature, or its heat is rejected to the environment, so the full $R\,I^2$ reduces useful power. Here the dissipation occurs within a retained reservoir at $T_0 > T_T$: at most $R\,I^2(1 - T_T/T_0)$ can be recovered, while $R\,I^2\,T_T/T_0$ is the unavoidable thermodynamic heat rejection. As $T_0$ approaches $T_T$, or if dissipation occurs outside the retained absorber, the recoverable fraction vanishes and the conventional full $R\,I^2$ loss is recovered. This thermodynamic heat rejection is not parasitic contact heat leakage.

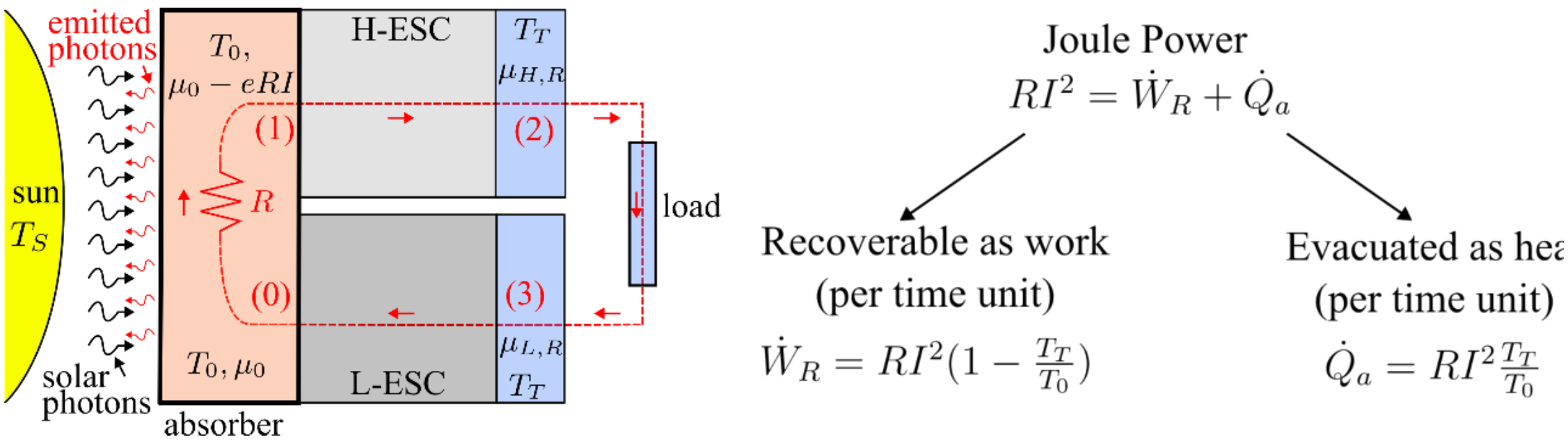

Fig. 2. Thermodynamic scheme of the non-ideal HCSC/HASC with finite absorber resistance. The diagram separates the Joule power $R\,I^2$ generated within the hot absorber into the maximum recoverable contribution $R\,I^2(1 - T_T/T_0)$ and the unavoidable heat $R\,I^2\,T_T/T_0$ rejected at the cold reservoir; the two branches sum to $R\,I^2$. This bookkeeping assumes ideal ESCs and no independent parasitic contact heat leakage. Labels in the absorber and ESC terminals indicate the temperature ($T$) and electrochemical potential of the electrons ($\mu$) at different points.

## 4. The hot absorber solar cell (HASC)

The "hot absorber" interpretation of hot-carrier conversion—i.e., allowing carrier–phonon equilibration while thermally insulating the absorber from the surroundings—has been pointed out earlier. [16–20] In this manuscript we use the term hot absorber solar cell (HASC) purely to label this thermally insulated-absorber operating regime and to distinguish it from the conventional hot-carrier narrative [21] that emphasizes suppressed carrier-to-phonon thermalization. As mentioned in the previous section, another common assumption for the ideal HCSC absorber is that electron-phonon interactions within the absorber are forbidden therefore ensuring that electron energy is not dissipated as heat within the absorber. As a result, except for the energy lost through emitted photons, all the energy captured from the photons in the absorber is efficiently transferred to the ESCs. Many hot-carrier proposals aim to reduce carrier cooling relative to the lattice, for example by engineering "phonon bottlenecks" (including hot-phonon scenarios in which emitted optical phonons are re-absorbed, and scenarios that suppress optical-to-acoustic decay such as Klemens-type channels). [21] These approaches attempt to maintain carriers hot while keeping the lattice comparatively cold. [22].

In contrast, the perspective adopted here is different: we do not assume suppressed electron–phonon interaction. In the HASC interpretation the absorber (carriers + lattice) is allowed to equilibrate internally and can become hot; the key idealization is instead that net non-radiative heat leakage from the absorber to the external environment is minimized (e.g., via vacuum isolation against conduction/convection), so that the absorber behaves as the hot reservoir in the same energy-conservation balance used for the ideal HCSC. Before explaining this model, it is necessary to review the properties that are demanded for the surface of an ideal HCSC absorber, properties not often made explicit in literature.

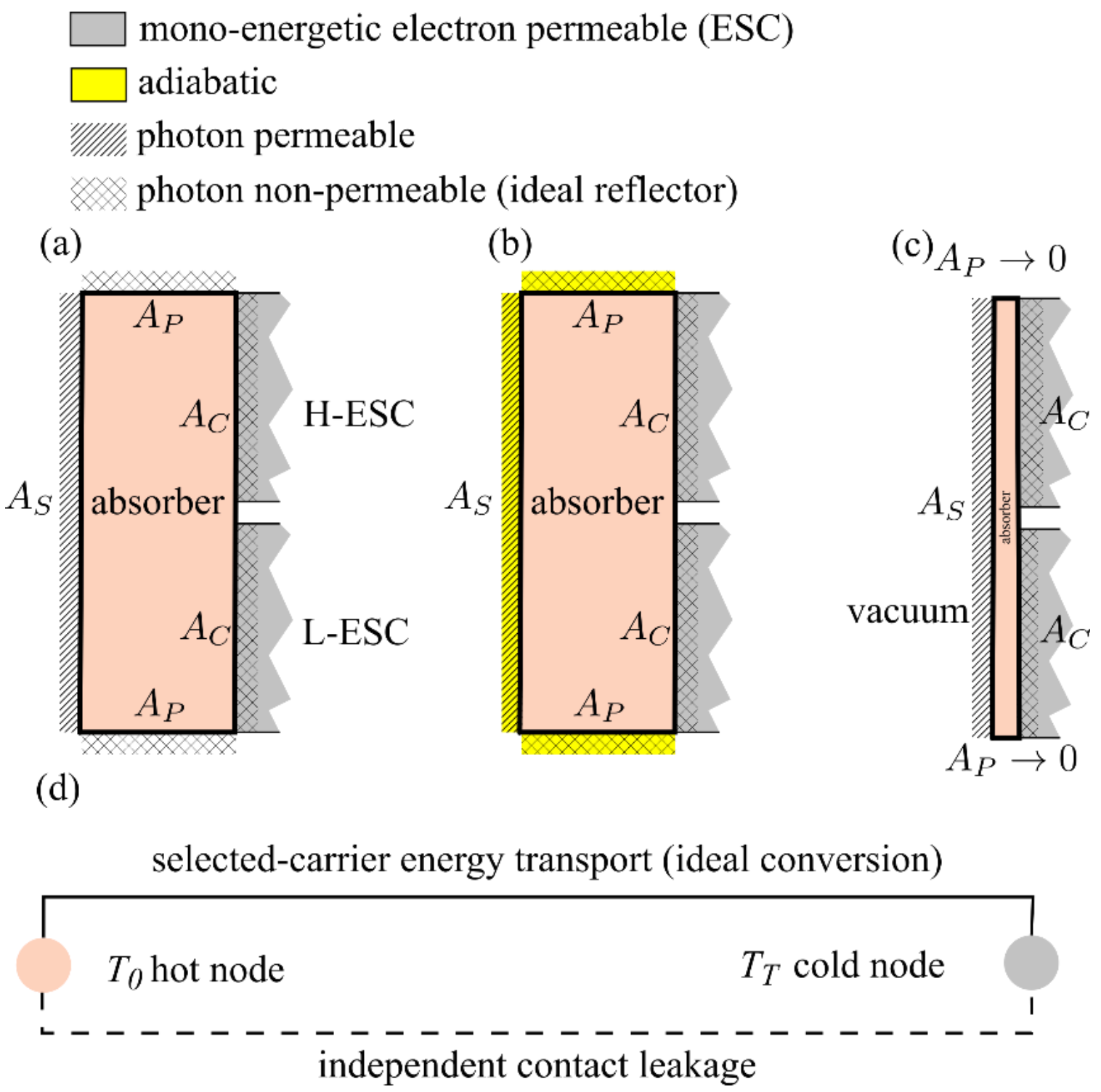

Fig. 3. (a) Surface properties of an ideal HCSC absorber; (b) HASC free-surface boundaries, where yellow denotes adiabaticity to nonradiative heat; (c) Illustrative ideal-contact geometry for a HASC. (d) Thermal-circuit representation with a solid selected-carrier conversion branch and a parallel dashed leakage branch, $\dot{Q}_{leak,ESC} = g_{ESC}(T_0 - T_T)$. The ideal model requires $g_{ESC} = 0$; the relaxed case has $g_{ESC} > 0$.

The surface of a HCSC absorber can be sub-divided in three parts: the surface that receives photons from the sun ($A_S$); the surface of absorber that is contact with the ESCs ($2A_C$), where the areas of the ESCs will be assumed to be the same for simplicity and equal each to $A_C$; and the rest of the surface, $A_P$, which typically corresponds to the area associated to the perimeter of the cell. The area between the ESCs will be assumed negligible (Fig. 3,a).

In the case of the ideal HCSC (the one in which electrons are not allowed to interact with phonons), using thermodynamic terminology, the surface $A_S$ is said to be permeable to photons, since it is the one that receives photons from the sun and emits also photons as black-body radiation at temperature $T_0$. It is rarely mentioned, however, that, in order the model of the ideal HCSC to be accurate, the surface $A_P$ has not to emit photons, that is, must be impermeable to photons or, in a more common language, must be a perfect mirror. The surface $A_C$, where the ESCs are attached to the absorber, must be permeable to mono-energetic electrons but has also to be totally reflective (impermeable) to photons. There is, though, some flexibility in this respect since, for example, if the ESCs were transparent and perform as an optical guide, the reflector could be located also at their other end, that is, at the cold side of the ESCs. These assumptions, related to the properties of the surface of the ideal absorber, are the ones that allow strictly formulating Eq. (1) which is at the core of the energy conservation model.

However, the energy conservation model can still be preserved if we lift the prohibition to electrons to interact with phonons, but we change some properties of the absorber surface. These new properties are the ones that lead to the HASC model (Fig. 3b). In this model, the surface $A_S$ is, of course, still considered permeable to photons, as in the ideal HCSC model, but we now demand it is also adiabatic (meaning it does not allow entropy, other than the one transported by the photons, to pass through). This could be achieved in practice, for example, by placing the surface $A_S$ in vacuum. As for the surface $A_P$, we now demand it is also adiabatic what can also be achieved by placing the absorber in vacuum. As for the surface $A_C$, the ideal HASC boundary instead requires both monoenergetic carrier exchange and zero independent parasitic contact conductance, $g_{ESC} = 0$. Vacuum surrounding the absorber suppresses gas conduction and convection at the free surfaces $A_S$ and $A_P$, but it does not eliminate solid-state heat conduction through electrically connected ESCs. In any case, the impact of the perimeter in the performance of both cells can be minimized by assuming thin absorbers in which the ratio $A_P/A_S \rightarrow 0$ (Fig. 3c). In consequence, Eq. (1) is also a mathematical model valid for the HASC and therefore, both the HCSC and the HASC exhibit the same limiting efficiency.

Ideal monoenergetic transport locks the desired carrier-energy flux to the selected particle current, but energy selectivity alone does not guarantee zero phonon conductance. Electrical exchange and low thermal conductance are not thermodynamically contradictory, although their simultaneous realization is a strong materials and device constraint (Refs. [6,21,23]). Possible design directions include vacuum-gap tunneling or thermionic energy filters, double-barrier resonant-tunneling structures with phonon suppression, and suspended or low-cross-section contacts. These are possible approaches, not demonstrated solutions across 1500–5000 K.

The advantage for the HASC, in our view, is that placing the absorber in vacuum seems more feasible in practice than demanding, for example, an absorber in which electrons do not interact with phonons. An intuitive illustration of the physical distinction between the two models can be obtained by considering an observer who, upon "touching" the absorber and therefore interacting with its phonons would perceive its surface as hot in the case of the HASC, but as cold in the case of the HCSC.

From this point, since both cells are described by the same model, they also share a common optimization problem related to the absorber thickness when the absorber is non-ideal. The non-idealities that will be considered stem from two sources: photon radiative emission losses through the perimeter area and the finite electrical conductivity, which leads to a nonzero resistance $R$. The trade-off consists in the fact that, while a thinner absorber suppresses radiative losses, it increases $R$. On the other hand, a thicker absorber reduces $R$ at the cost of higher radiative losses. In any case, the reduction in absorber thickness will be assumed not to compromise total light absorption, either due to the use of light-trapping techniques or because strongly absorbing materials, such as 2D semiconductors, are considered. A model combining both non-idealities is necessary to optimize the absorber thickness. This is done in the next section.

## 5. Non ideal hot absorber model

Before we start, we emphasize that this model, and the results derived from it, are valid both for the HCSC as for the HASC.

When radiative losses occur through the area $A_P$, Eq. (1) must be substituted by:

$$P_a - P_e - P_{e,P} - \dot{Q}_{leak,ESC} = \left(\frac{I}{e}\right)(\epsilon_H - \epsilon_L) \tag{23}$$

where, similarly to Eq. (3),

$$P_{e,P} = \hat{\sigma} T_0^4 A_P \tag{24}$$

is the power corresponding to the radiative photon emission through the area of the perimeter, $A_P$. It will be convenient to express the results scaled to the device unit area, $A_S$. Hence, dividing both sides of Eq. (23) by $A_S$, we obtain:

$$\hat{\sigma}T_S^4 - \hat{\sigma}T_0^4\left(1+\frac{A_P}{A_S}\right) - q_{leak,ESC} = \left(\frac{J}{e}\right)(\epsilon_H - \epsilon_L) \tag{25}$$

$$q_{leak,ESC} = \frac{\dot{Q}_{leak,ESC}}{A_S} = g_{ESC}(T_0 - T_T)$$

Here $g_{ESC} = (G_H + G_L)/A_S$ is the combined effective parasitic thermal conductance of the two contacts normalized to illuminated area, in W m$^{-2}$ K$^{-1}$.

$\dot{Q}_{leak,ESC}$ is a distinct parallel, current-independent thermal-conductance pathway that can remain finite at open circuit. By contrast, the selected-carrier energy rate (I/e)($\epsilon_H$ - $\epsilon_L$), the contact heat terms in Eqs. (9)–(12), and the absorber Joule-entropy heat in Eq. (16) all vanish when $I = 0$ and are not parasitic ESC leakage.

where $J = I/A_S$ becomes the electric current density calculated with respect to the total solar cell area. In terms of the more convenient current density, Eq. (19) and (20) can be rewritten as:

$$\mu_{H,R} = \{\mu_0 - e\hat{R}J\} + (\epsilon_H - \{\mu_0 - e\hat{R}J\})\left(1 - \frac{T_T}{T_0}\right) \tag{26}$$

$$\mu_{L,R} = \mu_0 + (\epsilon_L - \mu_0)\left(1 - \frac{T_T}{T_0}\right) \tag{27}$$

being

$$\hat{R} = RA_S \tag{28}$$

the resistance multiplied by the cell area.

As in the ideal case, Eqs. (25) to (27) allow calculating the current density vs. voltage characteristic of the solar cell with $T_0$ as parameter for a given value of $\hat{R}$. To calculate $\hat{R}$ we will assume the simple model for the electron current-density lines illustrated in Fig. 4. Mathematically, this model is described by the following set of equations:

$$\boldsymbol{j_n} = \begin{cases} \dfrac{I}{\frac{L}{2}\times L}\boldsymbol{j} & -\frac{L}{2} < x < 0;\ 0 < y < -\frac{W}{L/2}x;\ \ 0 < z < L \\ \dfrac{I}{W \times L}\boldsymbol{i} & -\frac{L}{2} < x < 0;\ -\frac{W}{L/2}x < \ y < W; 0 < z < L \\ \dfrac{I}{W \times L}\boldsymbol{i} & 0 < x < \frac{L}{2};\ -\frac{W}{L/2}x < \ y < W; 0 < z < L \\ -\dfrac{I}{\frac{L}{2}\times L}\boldsymbol{j} & 0 < x < \frac{L}{2};\ 0 < y < \frac{W}{L/2}x;\ \ 0 < z < L \end{cases} \tag{29}$$

Introduced into Eq. (15), they yield:

$$R = \frac{2\rho W}{L^2} + \frac{\rho}{2\,W} \tag{30}$$

and therefore,

$$\hat{R} = RA_S = RL^2 = 2\rho W + \frac{\rho L^2}{2\,W} \tag{31}$$

The first summand corresponds to the contribution to the resistance of the current flow in the $\boldsymbol{j}$ direction**,** degrading the series resistance as expected as $W$ increases. The second summand corresponds to the current flow in the $\boldsymbol{i}$ direction and increases if $W$ decreases, also as expected. The cell area, $L^2$, contributes through the second term.

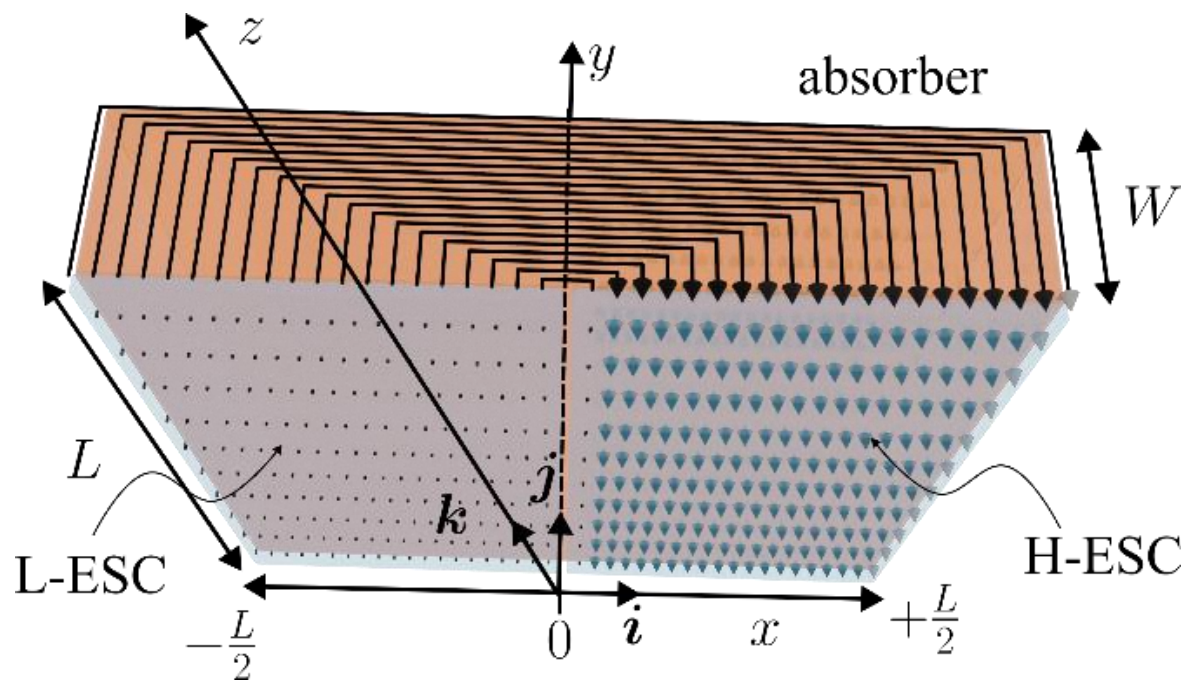

Fig. 4. Graphical representation of the electron current flow lines assumed in the calculation of the electrical resistance of the absorber. Current-density lines enter the absorber through the L-ESC, undergoes curvature within the material, and exit via the H-ESC.

The first term can be neglected if:

$$2\rho W \ll \frac{\rho L^2}{2\,W} \rightarrow \frac{L}{W} \gg 2 \tag{32}$$

We will assume that the aspect ratio determined by Eq. (32) is a reasonable approximation and consider, from now on, that:

$$\hat{R} \approx \frac{\rho L^2}{2\,W} = \frac{1}{2}\rho_\square L^2 \tag{33}$$

where

$$\rho_\square = \frac{\rho}{W} \tag{34}$$

is the sheet resistance of the absorber. This will also simplify the description of the results since, with this simplification, the electric resistance of the absorber results will depend only on $\rho_\square L^2$.

Finally, in a more general case, an actual solar cell can consist of several of the unit elements illustrated in Fig. 4 electrically connected in parallel, as sketched in Fig. 5, representing an array of $N \times N$ of these unit cells. We will call this configuration *cell array*. The total current circulating through the cell array, $I_T$, will be:

$$I_T = (N \times N)I \tag{35}$$

where $I$ is the current circulating through *each* unit cell. For the cell array, the cell area and perimeter area are given by:

$$A_S = (N \times N)L^2 \tag{36}$$

$$A_P = 4NWL \tag{37}$$

The new cell array will exhibit a total resistance, $R_P$, defined by the analogous Eq. to (15):

$$\dot{S} = \int_{(N\times N)\Omega_0} \dot{s}\, d\Omega = R_P I_T^2 / T_0 \tag{38}$$

where the integral volume now corresponds to the total volume of the $N \times N$ times the volume of the unit cell. Since the integral in the volume of each unit cell is given by Eq. (15) we have:

$$(N \times N)R\, I^2/T_0 = R_P I_T^2/T_0 \tag{39}$$

and therefore, using (35)

$$R_P = \frac{R}{N \times N} = \frac{1}{N \times N}\left(\frac{2\rho W}{L^2} + \frac{\rho}{2\,W}\right) \tag{40}$$

as otherwise expected from a parallel connection of identical resistors. Multiplied by the cell array area, we obtain:

$$\hat{R}_P = R_P A_S = R_P (N \times N) L^2 = 2\rho W + \frac{\rho L^2}{2\,W} \tag{41}$$

which equals $\hat{R}$ [Eq. (31)]. We now appreciate the advantage of formulating the operation of the solar cell in terms of the value of the resistance multiplied by the total cell area since $\hat{R}$ is invariant.

A final observation related to the cell array configuration concerns the radiation emitted from the perimeter of the cell. For the perimeter area between the cells, we notice that the photon emission from each cell balances the photon absorption due to radiative coupling between unit cells. Therefore, the internal perimeter area does not count for radiative losses but only the external perimeter area. Therefore, the array form of the generalized Eq. (25) remains valid with the explicit leakage term..

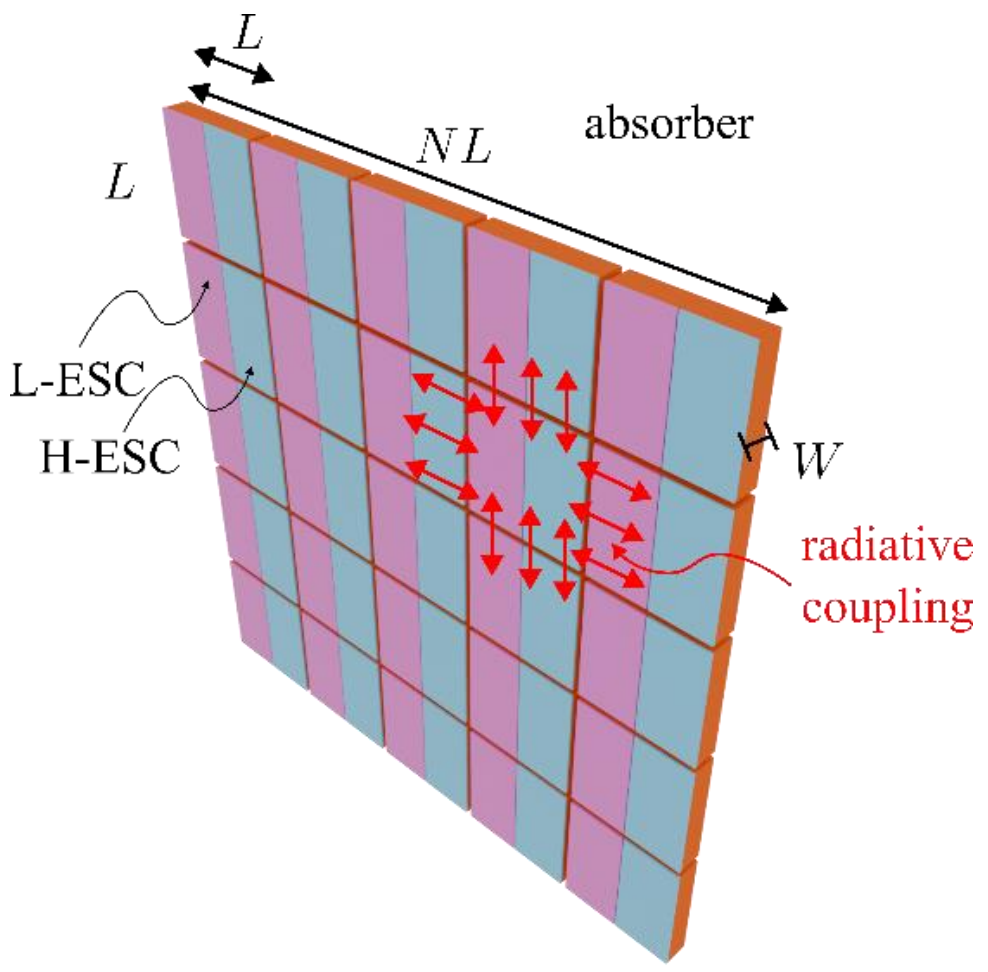

Fig. 5. Illustration of a parallel connection of unit HCSC elements. In this configuration, all the L-ESC are electrically connected with each other, and all the H-ESC are also connected with each other (the electrical connections are not shown). The balance of the net radiation emitted through the shared perimeter is zero because it cancels with the radiation absorbed.

## 6. Results

Once the model for the non-ideal absorber HCSC has been developed, we will explore next the impact that on the efficiency have: a) the existence of a perimeter with non-negligible area $A_P \neq 0$); b) the impact of finite mobility of the carriers in the absorber ($\rho_{\square} L^2 \neq 0$); c) the impact of heat recycling [Eq. (18)]; d) the impact of the ESC energy difference ($\epsilon_H - \epsilon_L$); and e) the impact of independent ESC heat leakage.

### *a) Impact of the surface-perimeter area ratio*

To explore the impact in the efficiency of the area of the perimeter, we will assume different cell area-perimeter aspect ratios defined by $A_S/A_P$, while we still assume that carriers through the absorber are transported without ohmic losses ($\hat{R} = 0$). The aspect ratio enters into the model through Eq. (25). Fig. 6 illustrates the results by plotting the current density-voltage characteristic of the HCSC (which we recall, are the same as for the HASC) for different aspect ratios. The conclusion of this calculation is that the efficiency of the HCSC tolerates relatively low aspects ratios since, for example, an aspect ratio in the order of the unity degrades the efficiency by only 2.5 % (from 85.3 % to 83.2 %). On the other hand, decreasing the aspect ratio has the apparent benefit of reducing the absorber temperature at which the maximum efficiency is obtained. This apparent benefit is misleading though because, for example, while a cell with $A_S/A_P$=0.1 operating at 1593.9 K exhibits a limiting efficiency of 76.7 %, a cell with aspect ratio $A_S/A_P$=100 operated at this same temperature would offer an efficiency of 80.8 %. Besides, for the general case of the cell array, we have:

$$\frac{A_S}{A_P} = \frac{(N \times N)L^2}{4NLW} = \frac{NL}{4W} \tag{42}$$

and, therefore, high $A_S/A_P$ ratios can be achieved by just increasing $N$.

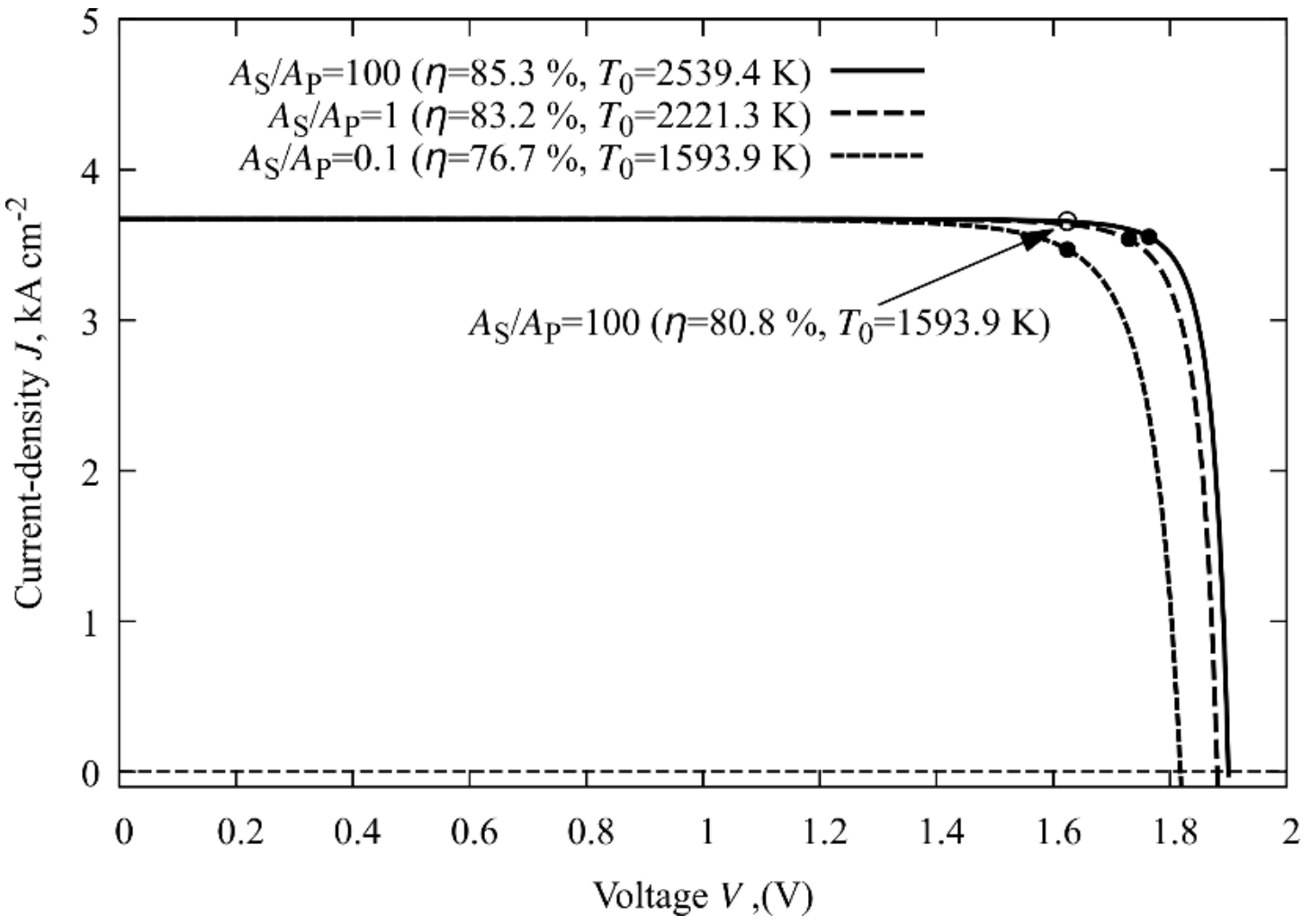

Fig. 6. Current-density vs voltage characteristics of a HCSC/HASC for different aspect ratios. The legend indicates also the corresponding efficiency and the absorber temperature at the maximum power point which is marked by a solid dot. The hollow dot indicates a point in the $A_S/A_P$=100 cell operating at the same temperature as the maximum power point of the $A_S/A_P$=0.1 cell. The absorber resistivity is zero and the ideal-contact benchmark $g_{ESC}$ = 0 is used. Other data: $\epsilon_H = 2$ eV; $\epsilon_L = 0$ eV; $T_T = 300$ K.

*b) Impact of finite electrical conductivity of the absorber.*

To explore the impact on the efficiency of the absorber electrical resistivity we will assume that at least an area aspect ratio $A_S/A_P = 100$ is achieved. We will also assume that Eq. (32) holds and, as discussed, $\hat{R}$ becomes only a function of the product $\rho_{\square}L^2$ [Eq. (33)]. Fig. 7a illustrates the impact on the efficiency of different values of $\rho_{\square}L^2$. To illustrate how to read this plot, we will take the case of $\rho_{\square}L^2 = 1 \times 10^{-7}$ Ωm² as an example. This value degrades the cell efficiency to 76.3 %. However, more problematic than this efficiency degradation is perhaps the fact that it comes associated with an increase in the absorber temperature up to the range of 3000 K. If we would assume, for example, that the absorber would be made of tungsten ($\rho \approx 10^{-6}$ Ωm at $T \approx 3000$ K) [24] in order to sustain this temperature range without melting, and if we would also assume an absorber thickness of $W = 10$ µm, this would imply $L \approx 1000$ $\mu$m which satisfy the previously assumed condition of $A_S/A_P$=100 for $N = 4$ as well as Eq. (32) since $\frac{L}{W} = 100$. Hence, an array of 4 × 4 unit cells would be needed to be integrated in parallel for this example.

Of course, higher values of $N$ are also allowed since they will not degrade cell efficiency. Although such scaling increases structural complexity, the example is intended only to illustrate how geometric scaling can, in principle, satisfy the combined constraints on $A_S/A_P$ and $\rho_\square L^2$; demonstrating a practical device would additionally require materials and contacts compatible with the resulting temperature range and with energy-selective transport. To relate the transport metric $\rho_\square L^2$ to familiar semiconductor values, note that (at 300 K) silicon spans resistivities from $\sim 4.6$ cm at $10^{15}\text{cm}^{-3}$doping down to $\sim 0.022\ \Omega\text{cm}$ at $10^{18}\ \text{cm}^{-3}$ (orders of magnitude variation). For an absorber thickness $W = 10\ \mu$m, these correspond to sheet resistances from $\rho_\square \approx 4.6 \times 10^3$ to $\approx 22\ \Omega/\square$, so achieving $\rho_\square L^2 \sim 10^{-7}\ \Omega\text{m}^2$ implies contact spacings $L$ ranging from a few micrometres (lightly doped) to $\sim 10^2\ \mu$m (heavily doped/degenerate). This illustrates that the high-efficiency regime requires either very high absorber conductivity, aggressive lateral scaling (small $L$), or both.

To dig in the interpretation of the results, we would now assume, instead, an absorber in the nm range, say $W = 10$ nm. With the same resistivity as tungsten, contacts should now be spaced $L \approx 30$ µm instead of 1000 $\mu$m to preserve that $\rho_\square L^2 = 1 \times 10^{-7}\ \Omega\text{m}^2$ and therefore, efficiency. Now $\frac{L}{W} = 3000$ so that the approximation in Eq. (32) is again valid, and $\frac{A_S}{A_P} = 750 > 100$ even for $N = 1$, no parallel integration of unit cells would be necessary. However, placing the ESCs 30 µm apart becomes technologically more demanding than 1000 $\mu$m.

*c) Impact of heat recycling.*

It is also illustrative to explore the impact on efficiency if the cell would not recycle into electrochemical energy part of the heat dissipated in the absorber. If this were the case, without the recycling factor $T_T/T_0$, Eq. (18) should instead be:

$$\dot{W} = \frac{I}{e}\left(\mu_{H,I} - \mu_{L,I}\right) - RI^2 = I\left[\frac{1}{e}\left(\mu_{H,I} - \mu_{L,I}\right) - RI\right] = IV \tag{43}$$

from which

$$V = \frac{1}{e}\left(\mu_{H,I} - \mu_{L,I}\right) - RI = \frac{1}{e}\left(\mu_{H,I} - \mu_{L,I}\right) - \hat{R}J \tag{44}$$

The current-density vs voltage characteristic obtained considering this output voltage instead of Eq. (21) is plotted in Fig. 7b for comparison. As expected, the impact of a higher effective series resistance translates into a lower fill factor and therefore lower efficiency. This impact is severe with the limiting efficiency decreasing, for example from 76.3 % to 24.0 % (for $\rho_\square$^$L^2 = 1 \times 10^{-7}\ \Omega\text{m}^2$) and the operation electron temperature rising to an impractically high value above 5000 K.

(a)

(b)

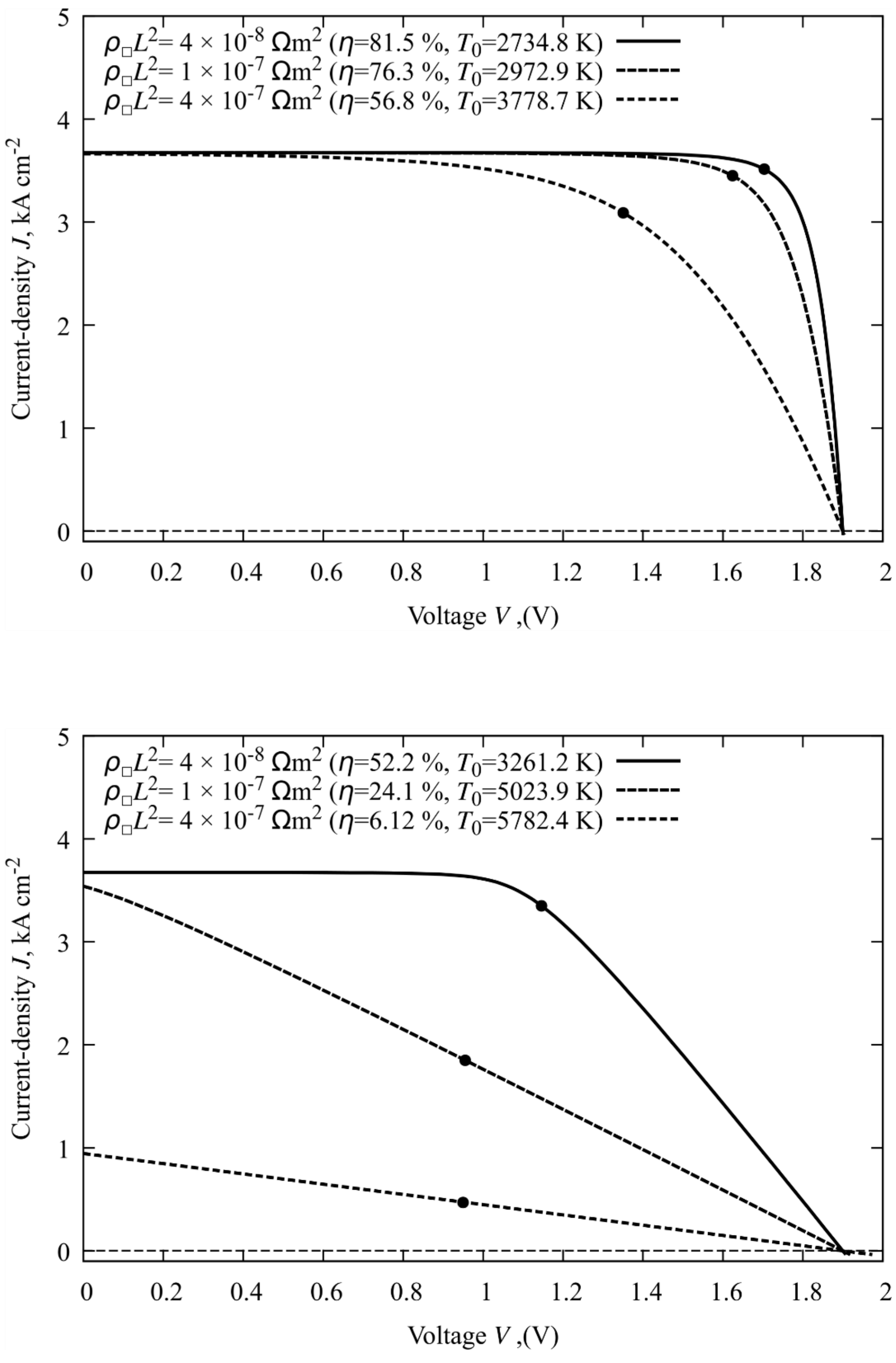

Fig. 7. (a) Current-density vs voltage characteristics of a HCSC/HASC for different values of the product $\rho_{\square}L^2$. The legend indicates also the corresponding efficiency and the absorber temperature at the maximum power point which is marked by a solid dot. (b) Counterfactual ordinary-series-loss/no-recycling comparison in which the full $R\ I^2$ is lost; this panel is not a model of ESC thermal leakage. Both panels use the ideal-contact benchmark $g_{ESC} = 0$. Other data: $\epsilon_H = 2$ eV; $\epsilon_L = 0$ eV; $T_T = 300$ K; $A_S/A_P$=100.

*d) Impact of energy contact difference.*

The dramatic impact of the series resistance on the limiting efficiency of the solar cell observed in the previous section suggests that devices should be designed towards higher operation voltages. To explore this approach, we will consider now the case of a solar cell in which $\epsilon_H -$ $\epsilon_L = 4$ eV (instead of 2 eV as considered in the previous examples). The results of the current-density vs voltage characteristic are plotted now in Fig. 8. Compared with the results for $\epsilon_H - \epsilon_L = 2$ eV shown previously in Fig. 7a, we observe, for example, that the same efficiency is obtained now for $\rho_\square$^$L^2 = 4 \times 10^{-7}$ Ωm² than in the case before for $\rho_\square$^$L^2 = 1 \times 10^{-7}$ Ωm². The reason why we get the same when we doble $\epsilon_H - \epsilon_L$ if we multiple by 4 the value of $\rho_\square$^$L^2$ is that, on the one hand by doubling $\epsilon_H - \epsilon_L$, Eq. (25) remains invariant if we multiply the current density, $J$, also by 2. On the other hand, since $\hat{R}$ impacts the output power per unit of area through $\hat{R}J^2$, this amount remains also invariant when $J$ is multiplied by 2 if $\hat{R}$ is divided by 4. This understanding can assist the future HCSC/HASC researcher in obtaining the required design parameters without the need for cumbersome calculations. As a final example, if we repeat the example of assuming a $W = 10$ $\mu$m thick absorber made of tungsten, now $L = 2000$ $\mu$m instead of $L = 1000$ $\mu$m and $N = 2$ instead of 4 would be necessary to reach $A_S/A_P$=100.

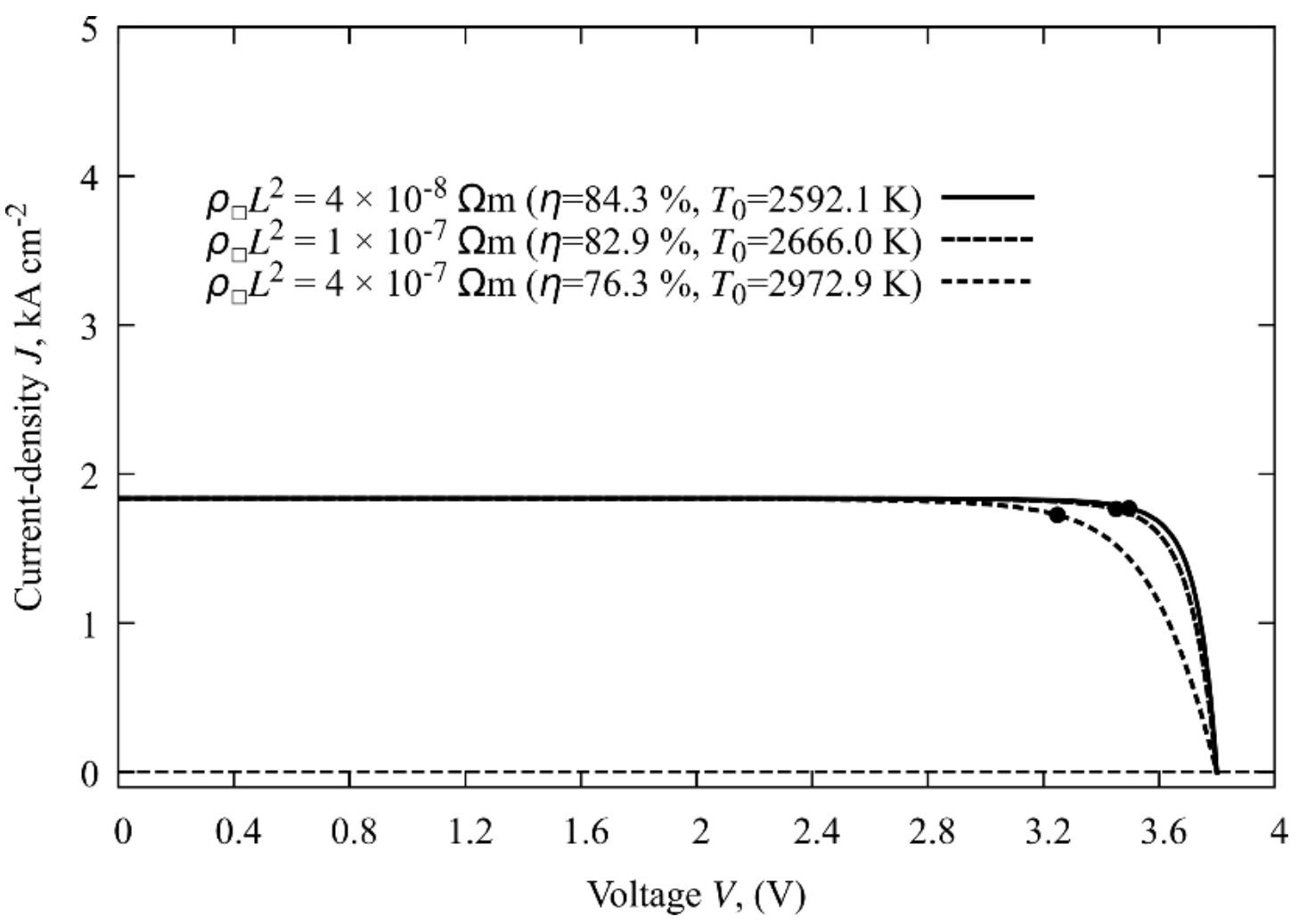

Fig. 8. Current-density vs voltage characteristics of a HCSC/HASC for different values of the product $\rho_\square L^2$. The legend indicates also the corresponding efficiency and the absorber temperature at the maximum power point which is marked by a solid dot. The ideal-contact benchmark $g_{ESC} = 0$ is used. Other data: $\epsilon_H = 4$ eV; $\epsilon_L = 0$ eV; $T_T = 300$ K; $A_S/A_P$=100.

### *e) Impact of heat leakage through the energy-selective contacts*

We now relax the ideal-contact limit by adding the parallel thermal-conductance term $q_{leak,ESC} = g_{ESC}(T_0 - T_T)$. For $T_S = 6000$ K, $T_T = 300$ K, $\epsilon_H - \epsilon_L = 2$ eV, $A_S/A_P = 100$, and $\rho_\square L^2 = 1.0 \times 10^{-7}$ Ω m², the maximum efficiency and $T_{0,MPP}$ are 76.26% and 2973 K at $g_{ESC} = 0$, 73.69% and 2854 K at 1 kW m$^{-2}$ K$^{-1}$, and 56.41% and 1962 K at 10 kW m$^{-2}$ K$^{-1}$.

For the ideal-resistance reference $\rho_\square L^2 = 0$, the corresponding values are 85.33% and 2539 K, 82.74% and 2396 K, and 66.63% and 1470 K for $g_{ESC} = 0$, 1, and 10 kW m$^{-2}$ K$^{-1}$, respectively. Figure 9 therefore quantifies how relaxing the $g_{ESC} = 0$ boundary condition changes the attainable performance.

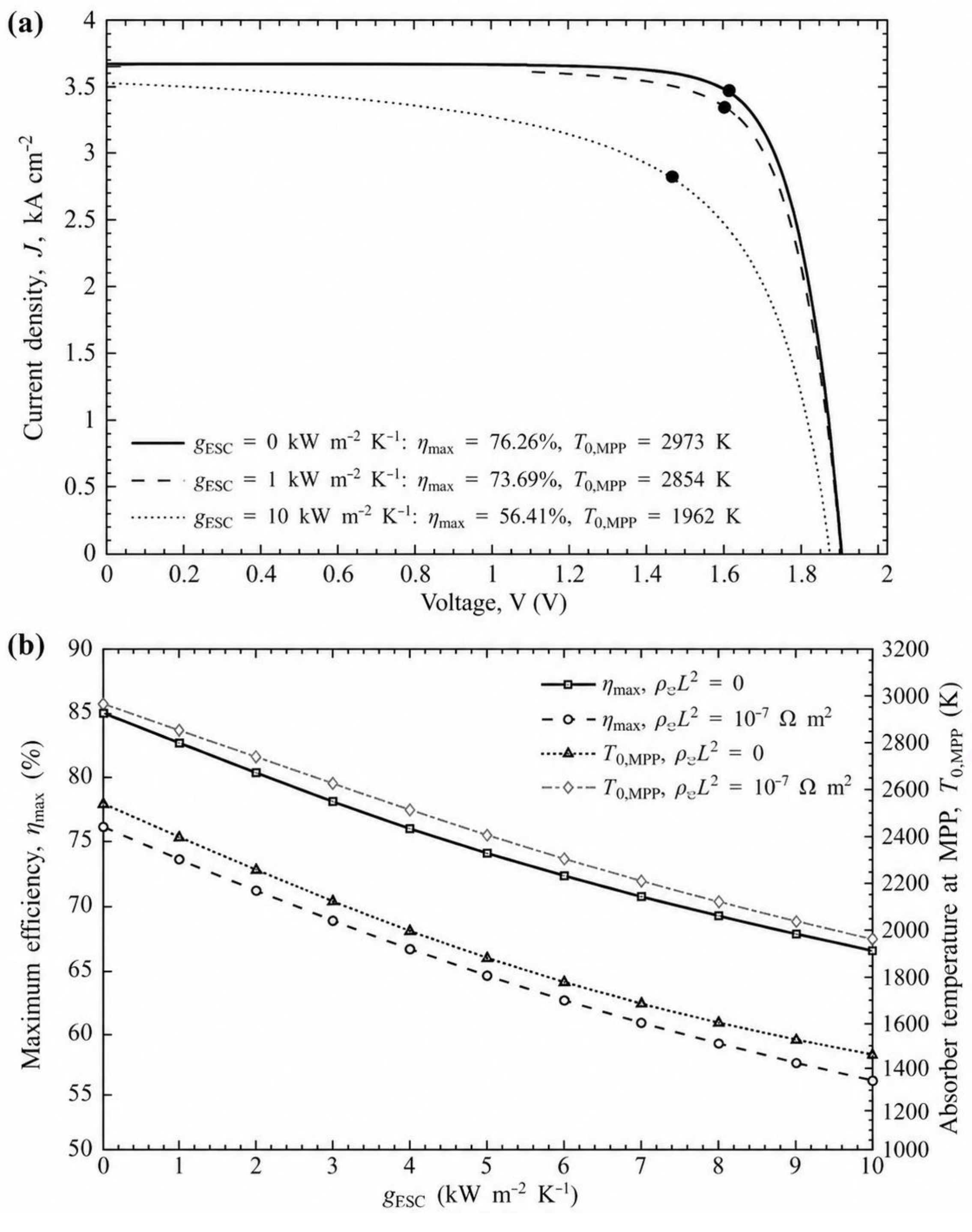

Fig. 9. Effect of parasitic heat leakage through the energy-selective contacts. (a) Current-density–voltage characteristics for several values of $g_{ESC}$ for $\rho$□$L^2 = 1.0 \times 10^{-7}$ Ω $m^2$; solid dots indicate maximum-power points. (b) Maximum efficiency and absorber temperature at the maximum-power point as functions of $g_{ESC}$ for an ideal-conductivity absorber and for $\rho$□$L^2 = 1.0 \times 10^{-7}$ Ω $m^2$. The limit $g_{ESC} = 0$ corresponds to the ideal contacts assumed in Figs. 6–8. Other data: $\epsilon_H = 2$ eV; $\epsilon_L = 0$ eV; $T_T = 300$ K; $A_S/A_P = 100$.

From a practical perspective, the temperatures $T_0$ that arise at the maximum-power point in Figs. 6–9 (typically 1500–5000 K in our parameter sweep) should be interpreted as the temperature of the hot reservoir in the underlying energy-conservation model. In the HASC interpretation, this reservoir includes both electronic and lattice degrees of freedom: we do not assume suppressed electron–phonon coupling. Instead, we assume that net non-radiative heat leakage from the absorber to the external environment is minimized (ideally, adiabatic boundaries), so that radiation and energy-selective carrier extraction dominate the exchange with the surroundings. This distinction is important: vacuum isolation does not prevent thermalization, but it can strongly reduce conduction and convection losses, leaving radiation as the main heat leak. [25] At such elevated temperatures, materials selection and encapsulation become central. [26] Candidate absorber platforms would likely require refractory chemistry (high melting point and low vapor pressure), for example refractory metals (e.g., W, Mo) or refractory carbides/nitrides (e.g., SiC, TiN, HfC/TaC) [27], potentially operated in an evacuated or inert-gas cavity analogous to high-temperature radiative elements. Vaporization is not negligible: tungsten vapor pressure has been measured in the 2600–3100 K range, [28] implying that long-term stability would require either lower operating temperatures, encapsulation, or replenishment strategies. Energy-selective contacts are an additional materials challenge; in the present work they are treated as ideal boundary conditions so as to isolate absorber-side constraints and to quantify the impact of Joule dissipation and perimeter radiation. The results should therefore be read as design targets—primarily $A_P/A_S$ and $\rho_\square L^2$ - rather than as a specific device prescription. At these temperatures, ESC band alignment, energy-window width, thermal conductivity, chemical stability, thermal expansion, and interfacial stability all change and must be assessed together with the refractory absorber chemistry, melting temperature, vapor pressure, and optical/electrical stability. No demonstrated absorber/contact stack is presently known to combine the required energy selectivity, electrical conductance, very low parasitic thermal conductance, and stability across this temperature range. The calculated efficiencies are thermodynamic limits and design targets rather than practical device-efficiency projections.

Importantly, the model indicates that increasing the ESC energy separation (raising operating voltage) and reducing $\rho_\square L^2$ can lower the required $T_0$ at a given efficiency, enabling a continuous trade-off between peak efficiency and materials feasibility.

## 7. Conclusions

A hot carrier solar cell (HCSC) and a hot absorber (HASC) solar cell share the same physical model and, therefore, the same limiting efficiency. The advantage of the second is that the absorber material does not need to fulfill, for example, the demanding condition by which electrons are not allowed to interact with phonons or must be quickly extracted before they thermalize. Instead, the absorber must behave as a thermally retained hot reservoir. Vacuum suppresses gas conduction and convection at the free surfaces, but the ideal HASC boundary also requires monoenergetic carrier exchange with zero independent parasitic contact conductance, $g_{ESC} = 0$.

The impact on the efficiency of non-ideal absorbers, in the sense of exhibiting finite electrical conductivity and radiative photon emission through the perimeter has been analyzed.

Concerning the impact of the radiative emission from the perimeter, it has been found that the HCSC/HASC is relatively tolerant to high area aspect ratios since, for example, a low ratio degrades the limiting efficiency by 2.5 %. In any case, the area aspect ratio can be increased by connecting unit cells in parallel without affecting the total resistance multiplied by the area of the cell, $\hat{R}$. In this respect, a model has also been developed that allows to relate the efficiency of the HCSC/HASC only to $\rho_\square$^$L^2$, being $\rho_\square$ the sheet resistance of the absorber and $L$ the length of the side of solar cell (and also the distance between contacts). In these conditions of finite electrical conductivity, it has also been found that the HCSC/HASC is capable of recycling into electrochemical work, within the limits of Carnot efficiency, part of the heat dissipated in the absorber because of Joule effect associated to the finite electrical conductivity. This allows obtaining the same limiting efficiency with thinner absorbers when comparing to the case in which this heat would not be recycled.

Finally, increasing the energy difference of the selective contacts produces higher output voltage HCSC/HASC at their maximum operation point what relaxes the need for low $\rho_\square$^$L^2$ so that if for example, the value of $\epsilon_H - \epsilon_L$ is multiplied by $x$, $\rho_\square$^$L^2$ can be multiplied by $x^2$ to obtain the same limiting efficiency.

By combining the results of this study on non-ideal hot carrier absorbers with our previous work on non-ideal energy selective contacts (Ref [6]), we conclude that a complete set of realistic materials for a functional hot carrier solar cell has yet to be identified. Regarding contacts, our previous research indicates that materials with equivalent thermoelectric $ZT$ factors of approximately 8 may be required; however, state-of-the-art thermoelectric materials such as GeTe and SnSe currently exhibit $ZT$ values only between 2 and 2.7 [29]. Concerning the absorber, two primary strategies exist: the first, exemplified by the HCSC absorbers in this work, seeks to suppress electron-phonon interaction to allow electrons to maintain high temperatures independently of the lattice. To date, no material has convincingly demonstrated the required suppression, which must also be preserved during transport through the energy selective contacts to prevent thermal energy from dissipating into the contact lattice before conversion into electrochemical energy. The second approach, represented by the HASC absorbers studied here, permits electron-phonon interaction in the absorber at the cost of equilibrating the lattice and electron temperatures. Operating at approximately 2500 K, this requires refractory materials which, despite their existence, present significant implementation challenges, as the selective contacts must also withstand these extreme operating temperatures. The present results should therefore be interpreted primarily as thermodynamic design targets and constraints, rather than as a prescription for an immediately realizable device.

The highest efficiencies correspond to the ideal-contact limit $g_{ESC} = 0$. Finite ESC leakage can substantially reduce both efficiency and $T_0$, and the compatibility of electrical exchange with thermal isolation is therefore an explicit ideal boundary condition and a major practical constraint. The Joule-heat recycling conclusion remains valid under the specified hot-reservoir, reversible-contact conditions, but its magnitude is reduced when bypass leakage changes $T_0$ and $I$.

## 8. Acknowledgments

The authors acknowledge financial support from the MAD2D-CM-UPM Project, funded by Comunidad de Madrid, the Recovery, Transformation and Resilience Plan (PRTR), and NextGenerationEU from the European Union; from grant TED2021-129694B-C21 (DEFY-CO2) funded by MCIN/AEI/10.13039/501100011033 and the PRTR; from grant TEC-2024/ECO-72 ("Por una Tecnología Amigable de Bajo Impacto Ambiental, 4EVERPV-CM") funded by Comunidad de Madrid in the 2024 call; from grant PID2021-124193OB-C21 (PVBooster Project) and PID2022-142425OA-I00 (COMIC) funded by MCIN/AEI/10.13039/501100011033. I. Durán acknowledges the financial support from the Ministry of Science, Innovation and Universities of Spain through the FPU fellowship program (grant number FPU21/03391).

## 9. Conflict of Interest

The authors declare no conflict of interest.